\documentclass[5p,twocolumn]{elsarticle}
\usepackage{graphicx,latexsym}
\usepackage{dcolumn}
\usepackage{amssymb,amsmath,bm}
\usepackage{subfigure}
\usepackage{braket}
\usepackage{siunitx}
\usepackage{array}
\usepackage{float}
\usepackage[nopar]{lipsum}
\usepackage{caption}
\usepackage{multirow}
\usepackage{bigstrut}

\biboptions{sort&compress}

\usepackage[super]{nth}

\newcommand{\angstrom}{\text{\normalfont\AA}}
\usepackage{hyperref}
\hypersetup{
    pdfnewwindow=true,       % links in new window
    colorlinks=true,         % false: boxed links; true: colored links
    linkcolor=blue,          % color of internal links
    citecolor=blue,          % color of links to bibliography
    filecolor=magenta,       % color of file links
    urlcolor=black           % color of external links
}

\usepackage[normalem]{ulem}

\def\sec#1{Sec.\ \ref{#1}}

\def\fig#1{Fig.\ \ref{#1}}
\def\tab#1{Tab.\ \ref{#1}}

\journal{}

\begin{document}

\begin{frontmatter}

%-----------------------------------------------------------------

\title{Controlling  electronic, magnetic, thermal, and optical properties of boron-nitrogen codoped strontium oxide monolayer:\break Activation of optical transitions in the VL region}

\author[a1]{Nzar Rauf Abdullah}
\ead{nzar.r.abdullah@gmail.com}
\address[a1]{Division of Computational Nanoscience, Physics Department, College of Science,
             \\ University of Sulaimani, Sulaimani 46001, Kurdistan Region, Iraq}
%\address[a2]{Computer Engineering Department, College of Engineering,
%	\\ Komar University of Science and Technology, Sulaimani 46001, Kurdistan Region, Iraq}

\author[a3]{Hemn Gharib Hussein}
\address[a3]{Physics Department, College of Education,
	 University of Sulaimani, Kurdistan Region, Iraq}

\author[a4]{Vidar Gudmundsson}
\address[a4]{Science Institute, University of Iceland, Dunhaga 3, IS-107 Reykjavik, Iceland}

%----------------------------------------------------------------

\begin{abstract}

The electronic, thermal, magnetic and optical properties of BN-codoped strontium oxide (SrO) monolayers are studied taking into account the interaction effects between the B and the N dopant atoms. The indirect band gap of a pure two dimensional SrO is modified to a narrow direct band gap by tuning the B-N attractive interaction.
The B or N separately doped SrO leads to a metallic behavior, while a BN-codoped SrO has a semiconductor character. The strong B-N attractive interaction changes a non-magnetic SrO to a magnetic system and reduces its heat capacity. An ab initio molecular dynamics, AIMD, calculations are also utilized to check the ther­modynamic stability of the pure and BN-codoped SrO monolayers.
The band gap reduction of SrO increases the optical conductivity shifting the most intense peak from the Deep-UV to the visible light region. The red shifted optical conductivity emerges due to the B-N attractive interaction. In addition, both iso- and anisotropic characters are seen in the optical properties depending on the strength of the B-N attractive interaction.
It can thus be confirmed that the interaction effects of the BN-codopants can be used to control the properties of SrO monolayers for thermo- and opto-electronic devices.

\end{abstract}


\end{frontmatter}

\section{Introduction}

The interest in 2D materials has grown both theoretically and experimentally due to their unique physical properties and potentials for applications in areas such as energy storage, sensing, and photonics devices \cite{GUPTA201544, Glavin_2020, Dong2017}.
The origin of the unique physical and chemical properties of the 2D monolayer materials can be ascribed to the dimensionality effect and the modifications of their band structure \cite{Wei2017, Alaal_2016, ABDULLAH2020126578}.
The tunable band gap of monolayers can effectively influence most of their physical characters.

Many different 2D graphene-like materials have been modeled after the discovery of a graphene monolayer in 2004 \cite{doi:10.1126/science.1102896} such as graphene-like monolayer monoxides including MgO, BeO, CaO BaO, and SrO monolayers \cite{Luo17213, https://doi.org/10.1002/admi.201700688}. The electronic, the thermal, and the optical properties of some of these monolayers have been intensively studied and shown their great potential for applications.  

The investigation of the electronic properties of MgO monolayers show more fascinating character, than their bulk phase, such as the reduction of the band gap from $7.8$ eV to $3.1\text{-}4.2$ eV \cite{PhysRevB.95.144109}.
The optical conductivity of an MgO monolayer thus indicates its semiconductor property \cite{nourozi2019electronic, VANON2022106876}. In addition, it has been shown that an MgO monolayer exhibits a non-magnetic behavior in spin polarization.  If oxygen atom is substituted by nonmagnetic dopant atoms such as (B, C, N and F), magnetic half-metal, antiferromagnetic semiconductor, and nonmagnetic metal are obtained depending on the dopant atoms \cite{moghadam2018electronic}.

The BeO monolayer has also a large band gap making it a poor material for optoelectronic devices in a wide range of energy. If nonmagnetic atoms such as B or N are doped in BeO, it's band gap is tuned improving its optical properties \cite{ABDULLAH2022107102, C6RA04782C}. The inter-atomic interaction between the B and the N dopant atoms gives high thermoelectric and optical conductivity of a BN-codoped BeO monolayer \cite{ABDULLAH2022106409}. 

There are few studies on the SrO monolayer in which one of the study indicates that strontium oxide could be a good candidate for the growth of epitaxial perovskite oxides on semiconductors because the SrO acts as a buffer layer between the metal oxide reactive semiconductor layers \cite{wang2013atomic}. 
The Piezoelectric effect of the SrO monolayer has been studied and it was shown that it
is a strong candidate for atomically thin piezoelectric applications \cite{https://doi.org/10.1002/pssb.201600387}. 
Recently, the lattice constant, the bond length, and the binding energy and the thermodynamic stability of the SrO monolayer have been studied. The study indicates that the SrO monolayer is thermally stable up to $2000$ K \cite{doi:10.1073/pnas.1906510116}. The stability of the SrO monolayer is an interesting point that will instigate further studies. We thus investigate the electronic, the thermal, the magnetic, and the optical properties of pure and BN-codoped SrO monolayers in this work based on density functional theory.

The paper is organized as follows: The computational methods are presented in the \sec{methodology}. The calculated electrical, thermal, and optical properties are displayed in \sec{results}. The conclusions are drawn in \sec{conclusion}.

\section{Model and Computational tools}\label{methodology}

We consider a $2\times2$ super-cell monolayer consisting of pure SrO and BN-codoped SrO nanosheets.
First-principle calculations are performed based on density functional theory, DFT, within the framework of the projector augmented wave, PAW, method implemented in the Quantum espresso (QE) software \cite{Giannozzi_2009, giannozzi2017advanced}. The generalized gradient approximation, GGA, is used to approximate the exchange and the correlation terms with the assumption of the PBEsol functionals \cite{PhysRevB.54.11169}. 
A vacuum length of 20 $\angstrom$ is assumed to cancel the interlayer interaction in the periodic arrangement along the $z$-direction \cite{ABDULLAH2023140235, ABDULLAH2022115554}.
A cut-off plane-wave kinetic energy of $1100$~eV is selected in the plane-wave basis set, and 
the system is considered converged with the total energy in self-consistent field, SCF cycles set at $10^{-6}$~eV and the forces on individual atoms beeing less than $0.001$~eV/$\angstrom$ \cite{ABDULLAH2023116147}.
The band structure and the density of states (DOS) are obtained by utilizing the SCF, and non-self-consistent field (NSCF) calculations, respectively. In these calculations, we have used a Monkhorst-Pack grid of $18 \times 18 \times 1$ for the SCF and $100 \times 100 \times 1$ for the NSCF \cite{abdullah2021modulation}.
The optical properties of the monolayers are obtained using QE with the optical broadening of $0.1$~eV.
In addition, the thermal calculations are performed by the dmol$^3$ software based on a DFT
formalism \cite{doi:10.1063/1.458452}.
An ab initio molecular dynamics, AIMD, calculations have performed to see the possibility of  ther­modynamic stability of the monolayers. The calculations, done in the NVT ensemble, are utilized for $5$~ps with a time step of $1.0$~fs using the heat bath approach described by Nosé-Hoover \cite{doi:10.1063/1.463940}.

\section{Results}\label{results}

Different properties of B-, N-doped and BN-codoped SrO monolayers are calculated, such as the structural, the electronic, the magnetic, the thermal, and the optical characteristics. For the sake of comparison, the properties of pure SrO monolayer are also calculated. 
\begin{table*}[h]
	\centering
	\begin{center}
		\caption{\label{table_one} The average bond lengths of Sr-O, Sr-B, Sr-N, and B-O, the average lattice constant, $a$, the formation energy, E$_{f}$, and the band gap, E$_{\rm g}$.
			The unit of bond lengths is $\angstrom$.}
		\begin{tabular}{l|l|l|l|l|l|l|l}\hline
			Structure&  Sr-O      &  Sr-B     &  Sr-N    & B-O  & $a$ ($\angstrom$)   &  E$_f$  ($\rm eV$/atoms)  &  E$_{\rm g}$  ($\rm eV$)  \\ \hline
			SrO	     &  2.29      &  -        &  -       &  -   &  3.99  & 1.4  &   1.68    \\
			B-SrO    &  2.34      &  2.64     &  -       &  -   &  4.24  & 1.54 &    -      \\
			N-SrO    &  2.33      &  -        &  2.37    &  -   &  4.06  & 1.62 &    -      \\
			BN-SrO-I &  2.35      &  -        &  2.48    & 1.46 &  3.47  & 2.83 &  0.15     \\
			BN-SrO-II&  2.5       &  -        &  2.17    & 1.42 &  3.49  & 2.25 &  1.04     \\ \hline
	\end{tabular}	\end{center}
\end{table*}

\subsection{Structural properties}

\begin{figure}[htb]
	\centering
	\includegraphics[width=0.45\textwidth]{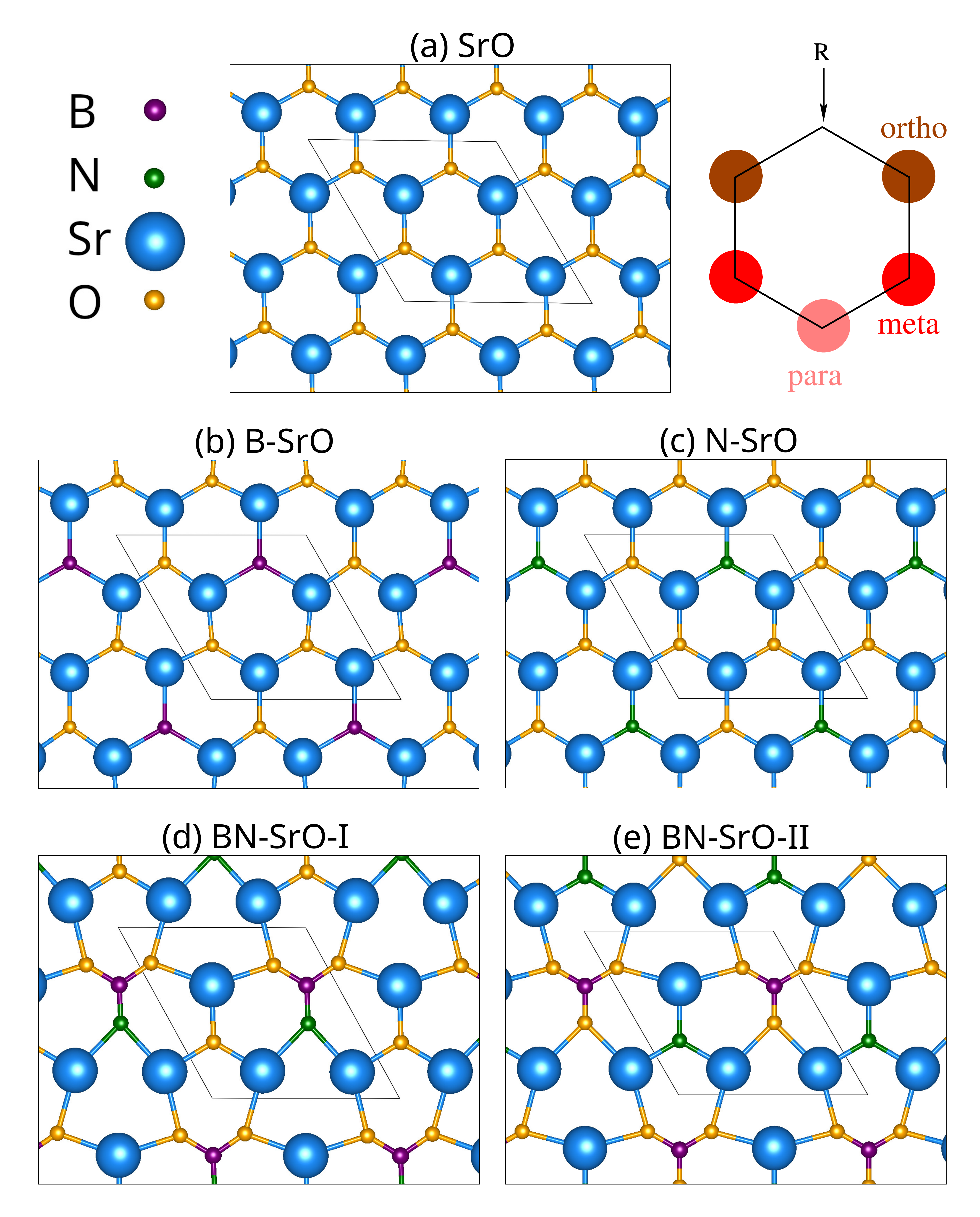}
	\caption{Atomic configuration of pure SrO (a), B-SrO (b), N-SrO (c), BN-SrO-I (d), and BN-SrO-II (e) monolayers. The hexagonal shape (top right) is a schematic diagram that shows atomic position of ortho, meta, and para.}
	\label{fig01}
\end{figure}

Several atomic configurations of SrO and BN-coped SrO monolayers are considered in this study which are shown in \fig{fig01}.
In addition to the pure SrO monolayer (a), the B doped SrO and N doped SrO identified as B-SrO (b) and N-SrO (c), respectively, are considered. In order to attain the most structurally stable monolayer, the B or the N atoms are substitutionally doped at the position of an O atom. In this case, no large deformation in the monolayer is seen as the atomic radius of an O atom is close to the atomic radius of an N and a B atom. 
We Note that if B or N atoms are doped at the Sr position of the hexagon, a deformation in the structure occurs leading to an unstable monolayer. 
We thus do not consider B or N substitutionally doped at the Sr atom positions. 

Two stable atomic configurations of the BN-codoped SrO are also assumed. First, the B atom is doped at an ortho-position and the N atom is doped at the meta-position identified as the BN-SrO-I monolayer (d). Second, a B atom is doped at the ortho-position while the N atom is doped at the meta-position (opposite position of meta) identified as the BN-SrO-II monolayer (e). These two cases leading to different distances between the B and tje N atoms in the SrO monolayer lead to different inter-atomic interactions.

The bond lengths, the lattice constant, the formation energy and the band gaps of these monolayers are presented in \tab{table_one}. 
The formation energy indicates a structural stability for all the considered monolayers as the values of the formation energy are not too high \cite{PhysRevB.92.115307}.
The Sr-O bond length and the lattice constant of a pure SrO monolayer are $2.29 \, \angstrom$ and $3.99 \, \angstrom$, respectively, in good agreement with the literature using DFT calculations \cite{doi:10.1073/pnas.1906510116}. The average Sr-O bond length and the lattice constant of the B-SrO and N-SrO monolayers are increased as is expected with the atomic radius of both B and N being larger than that of an O atom.

The bond length and the lattice constant in the BN-codoped SrO are not only affected by the atomic radius but the inter-atomic interactions between the dopant atoms play a role. 
In both the BN-SrO-I and the BN-SrO-II monolayers, the lattice constant is reduced comparing to the pure SrO.
This indicates that the monolayers are shrunk due to the attractive interaction between the B and the N dopant atoms. The attractive interaction between the B and the N atoms in BN-SrO-I is stronger than that of BN-SrO-II because the distance between the B and the N atom in BN-SrO-I is smaller than that of BN-SrO-II.
\begin{figure}[htb]
	\centering
    \includegraphics[width=0.23\textwidth]{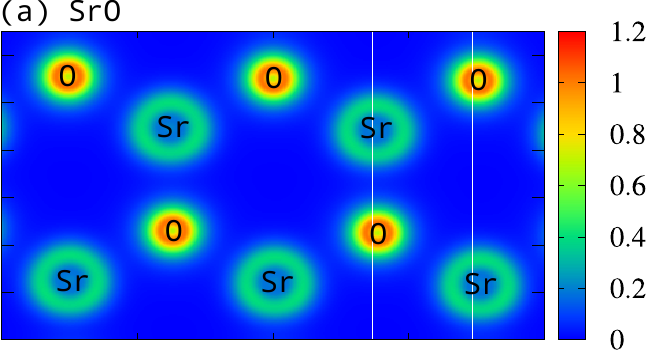}\\
	\includegraphics[width=0.23\textwidth]{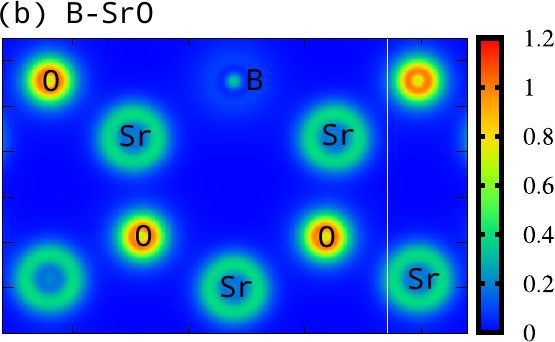}
	\includegraphics[width=0.23\textwidth]{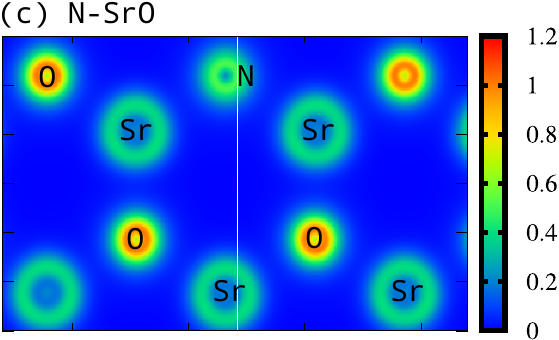}\\
	\includegraphics[width=0.23\textwidth]{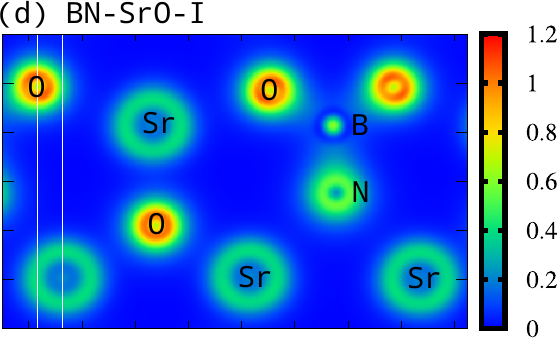}
	\includegraphics[width=0.23\textwidth]{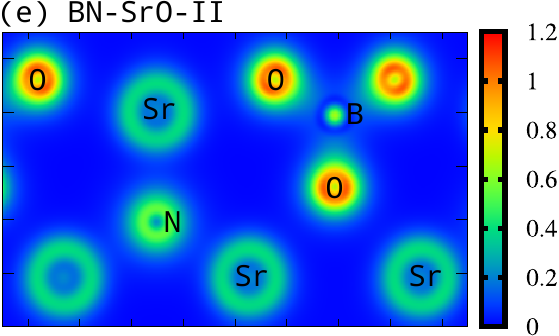}
	\caption{Electron density distribution of pure SrO (a), B doped SrO (b), N doped SrO (c), BN-SrO-I (d), and BN-SrO-II (e) monolayers.}
	\label{fig02}
\end{figure}

To see the effects of the atomic configurations of the dopant atoms, the electron density distribution, EDD, of the monolayers is displayed in \fig{fig02}. It is well known that the outer shell electrons of Sr, O, B, and N atoms are $5s^{2}$, $2s^2 2p^{4}$, $2s^2 2p^{1}$, and $2s^2 2p^{3}$, respectively.
The EDD for the pure SrO shown in (a) indicates that the electrons are mostly localized around the O atoms while a slightly electron localization is seen around the Sr atoms, which is due to transferring of electrons from the outer shell of the Sr atoms to the O atoms. 
In the case of B-SrO, one electron from the outer shell of a B atom transfers to the neighborhood atoms leading to a very weak electron distribution at the B atom, while the electrons are localized around the N atom in the case of an N-SrO monolayer due to the presence of the $5$ electrons outer shell of an N atom. 
The electron distribution of BN-SrO-I indicates an electron localization between the B and the N atoms due to the outer shell electron sharing between these two atoms forming a relatively strong covalent bond. This thus confirms a strong attractive interaction between the B and the N atoms. 
The attractive interaction between the B and the N atoms becomes weaker in the BN-SrO-II as the distance between these two atoms is larger there.

\subsection{Band structure and DOS}

\begin{figure}[htb]
	\centering
	\includegraphics[width=0.35\textwidth]{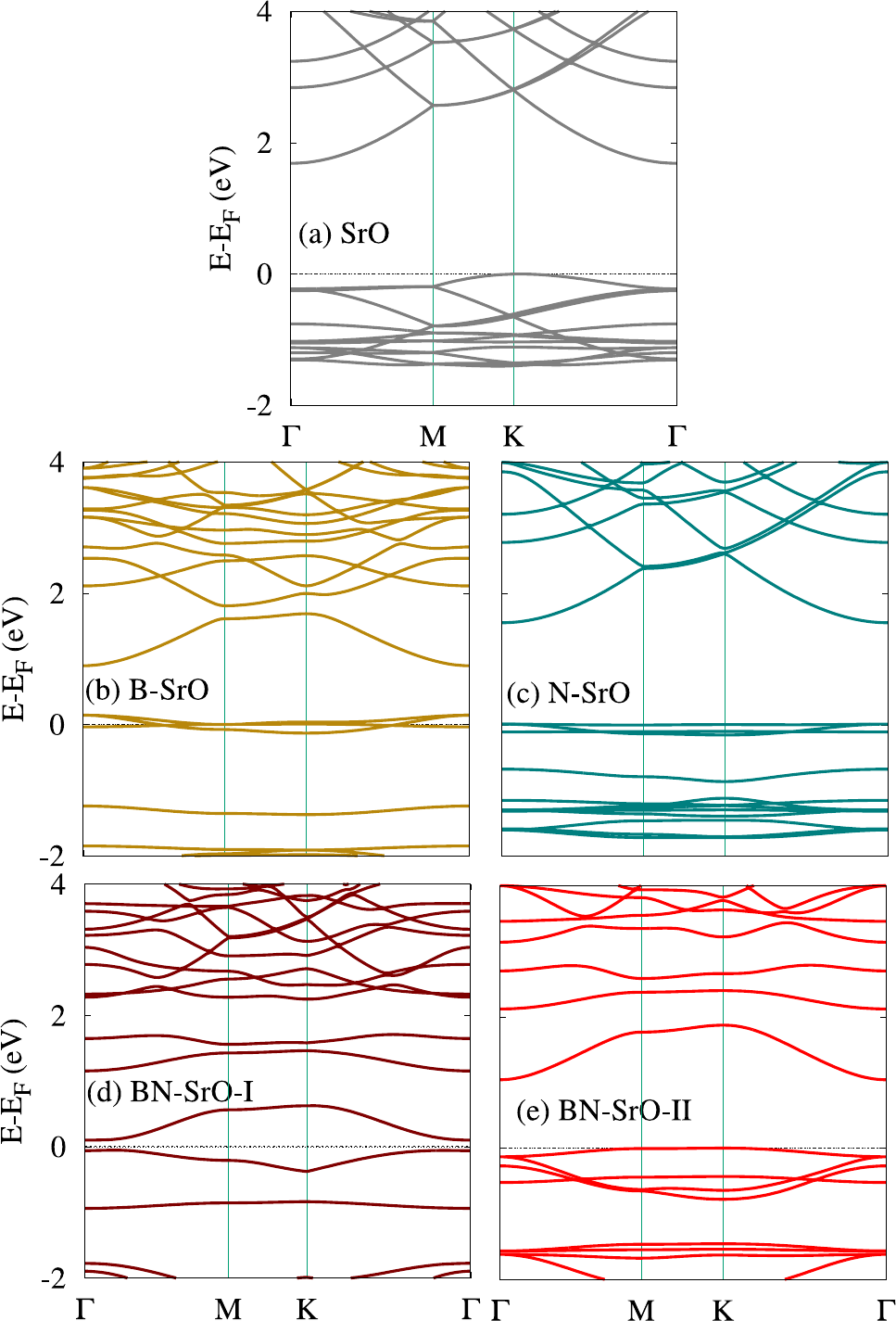}
	\caption{Band structure of pure SrO (a), B-SrO (b), N-SrO (c), BN-SrO-I (d), and BN-SrO-II (e). The energies are with respect to the Fermi level, and the Fermi energy is set to zero.}
	\label{fig03}
\end{figure}

The band structure, the density of states (DOS) and the partial density of states (PDOS) are computed here.
Figure \ref{fig03} shows the band structure of pure and doped SrO monolayers,
The pure SrO monolayer has an indirect gap of $1.68$~eV with the valence band maximum (VBM) 
and the conduction band minimum (CBM) located at the K point and the $\Gamma$ point, respectively. 
The band gap of pure SrO is in a good agreement with other DFT calculation \cite{doi:10.1073/pnas.1906510116}, and it displays that the SrO is a semiconductor material.

It is clearly found that the CBM (VBM) for undoped SrO moves below (above) the Fermi energy upon
doping by B (N).
If the B atom is doped in SrO, the CBM crosses the Fermi energy, while the N doped SrO introduces a shift up of the VBM crossing the Fermi energy. Both B-SrO and N-SrO monolayers have thus metallic property. 

\begin{figure}[htb]
	\centering
	\includegraphics[width=0.4\textwidth]{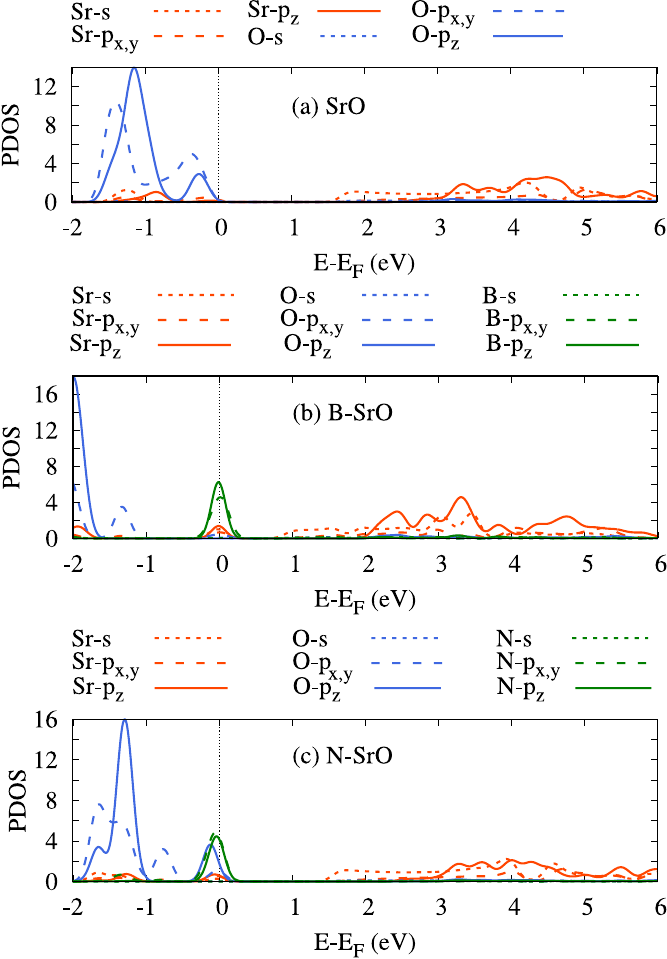}
	\caption{Partial density of state for SrO (a), B-SrO (b), and N-SrO (c) monolayers. The vertical dotted black is the location of Fermi energy in which the energy ($x$-axis) is with respect to the Fermi level, and the Fermi energy is set to zero.}
	\label{fig04}
\end{figure}

The PDOS can be used to understand the crossing feature of the bands. The PDOSs of pure SrO (a), B-doped SrO (b) and N-doped SrO (c) are shown in \fig{fig04}. The $x$- and $y$-components of the $p$-orbital of the O, Sr, B, and N atoms are symmetric for pure and doped SrO monolayers. 
This indicates the symmetry of monolayers along both the $x$- and the $y$-axis even in the cases of B- or N-dopant atoms.

Another interesting property of the PDOS of the pure SrO is that the $p$-orbitals of the O atoms are dominant in the valence band region, while the $s$- and the $p_z$-orbitals of the Sr atoms have a major contribution to the conduction bands. Especially the $s$-orbital of the Sr atoms contributes to the CBM.

In the case of B-SrO, the $p$-orbitals of the B atom generate the states at the Fermi energy and the B atom shifts the $s$ and $p$-states of the Sr atoms down to the Fermi energy level. This leads to crossing of the CBM and the Fermi energy in the band structure of B-SrO shown in \fig{fig03}(b). The question here is why the B atom shifts the $s$ and $p$-states of the Sr atoms down to the Fermi energy level?. To answer this, we refer to the atomic configuration of the corresponding atoms.  
The B atom ($2s^2 2p^1$) has one valence electron more than the Sr atom ($5s^2$). 
Therefore, doping by B acts like electron doping of SrO resulting in shifting of
the conduction band to accommodate the extra electron. We note that we compare the atomic configuration of the B atom with the Sr atom because the B atom is replaced with an O atom in the B-SrO monolayer resulting in a B-Sr bond, while no B-O bond is seen in the structure.

Opposite scenario can be applied for the case of the N-SrO monolayer. Instead, the N atom has one valence electron less than the O atom.  We can thus see the produced states at the Fermi energy due to the N atom (\fig{fig04}(c)), and shifting up the $p_z$ state of the O atom.
Therefore, doping by N acts like a hole doping of SrO resulting in shifting of the valence bands to accommodate the extra holes. Consequently, the VBM shifts up and crosses the Fermi energy.

\begin{figure*}[htb]
	\centering
	\includegraphics[width=0.8\textwidth]{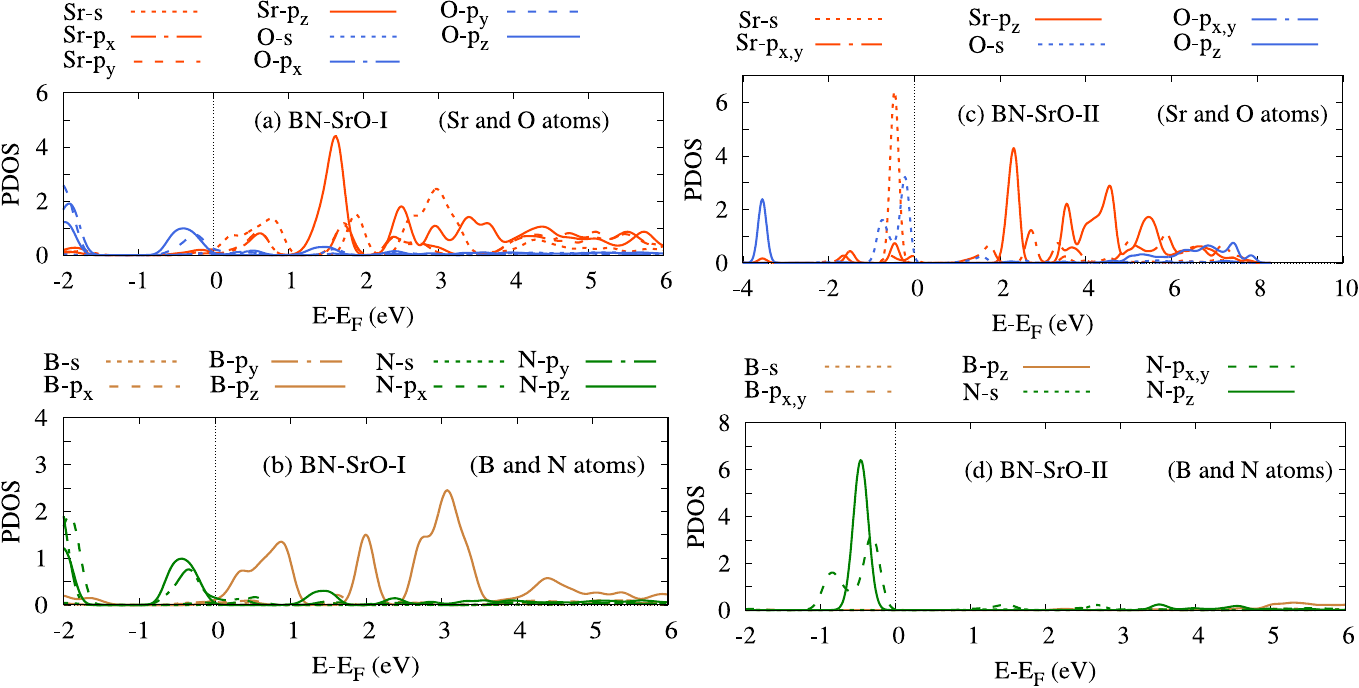}
	\caption{PDOS of BN-SrO-I (a,b) and BN-SrO-II (c,d) monolayers. The vertical dotted black is the location of Fermi energy in which the energy ($x$-axis) is with respect to the Fermi level, and the Fermi energy is set to zero.}
	\label{fig05}
\end{figure*}

The most interesting cases are the BN-codoped SrO monolayers for which the band structures are presented in \fig{fig03}(d,e), and their PDOS are shown in \fig{fig05} for BN-SrO-I (a,b) and BN-SrO-II (c,d).

\begin{figure}[htb]
	\centering
	\includegraphics[width=0.3\textwidth]{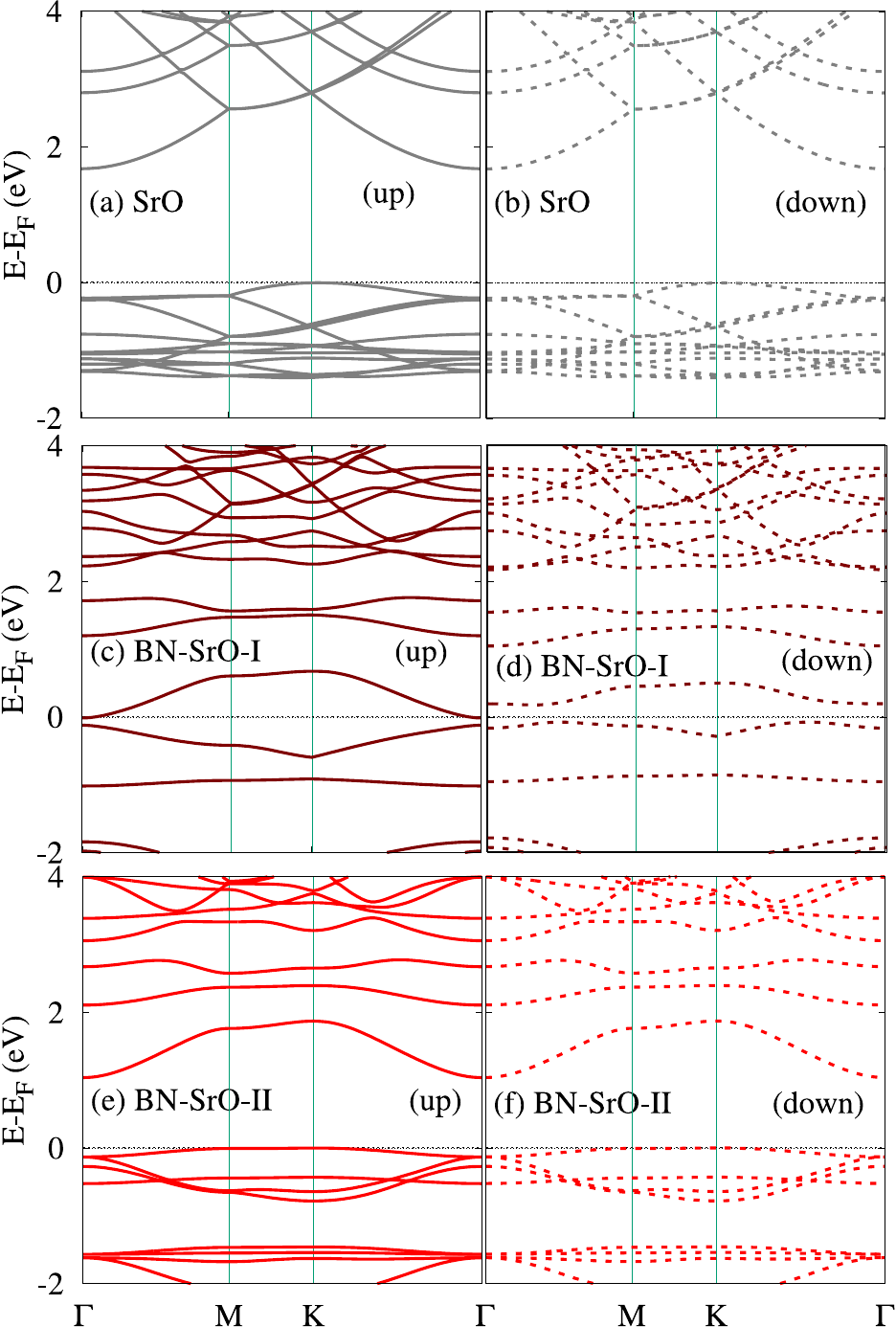}
	\caption{Band structure of pure SrO for spin up (a) and down (b), BN-SrO-I for spin up (c) and down (d), BN-SrO-II for spin up (e) and down (f). The energies are with respect to the Fermi level, and the Fermi energy is set to zero.}
	\label{fig06}
\end{figure}

In the cases of BN-codoped SrO, both B and N dopant atoms influence the crossing of the VBM and the CBM to the Fermi energy resulting a band gap reduction.
The strong attractive interaction between the B and the N dopant atoms in BN-SrO-I monolayer strongly reduces the band gap to $0.15$~eV (see \tab{table_one}), while a weak attractive interaction in the BN-SrO-II monolayer reduces the band gap to $1.04$~eV. This indicates that both BN-SrO-I and BN-SrO-II have semiconductor properties. One may be interested to see the PDOS of these two cases in \fig{fig06}. 
The first interesting point is that the strong attractive interaction in BN-SrO-I breaks the symmetry properties of the $p_x$ and the $p_y$ orbitals, while the symmetry character is still seen in the presence of the weak attractive interaction in BN-SrO-II. 

Another point to notice is that the O and the N atoms are dominant in the valence band region, while the Sr and the B atoms have the major contribution in the conduction band region. The strong attractive interaction in BN-SrO-I generates more localized states in the conduction band region via the $p_z$ states of the B atom, which forms more states in the conduction band region (see \fig{fig03}(d)) comparing to the band structure of BN-SrO-II.
This allows more transitions from the valence to the conduction band regions as is presented later.

The spin-dependent DFT calculations were done to check the magnetization of BN-codoped SrO monolayers. 
The band structures for both spin components (up and down) are calculated as is demonstrated in  
\fig{fig06}. It seems that the band structures of spin up and down for pure SrO and BN-SrO-II are exactly the same indicating a non-magnetic semiconductor character. On the other hand, the band structure of BN-SrO-I is different for spin up and down, especially around the Fermi energy displaying a magnetic material behavior of BN-SrO-I. We note that the BN-SrO-I has semiconductor property for both spin components. The asymmetric behavior of the spin up and down can be referred to the strong attractive interaction between the B and the N atoms in the structure. 

\subsection{Thermal properties}

\begin{figure}[htb]
	\centering
	\includegraphics[width=0.5\textwidth]{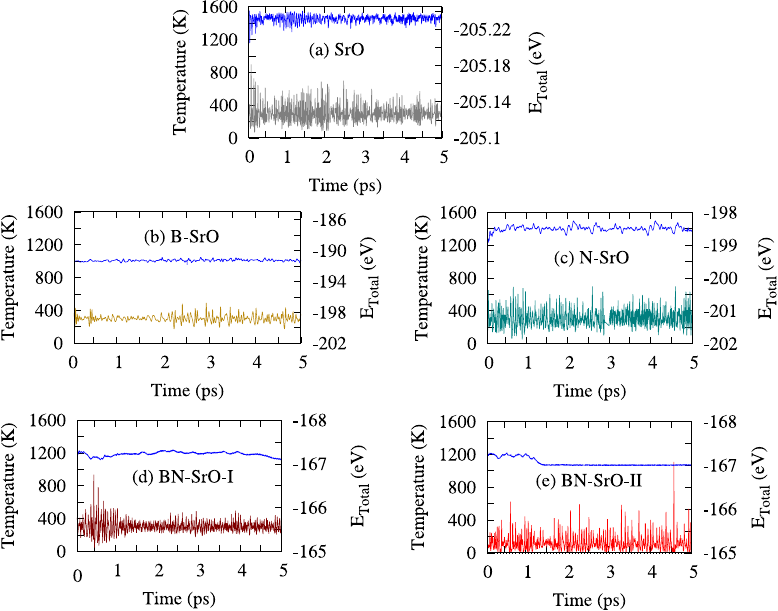}
	\caption{Temperature versus the AIMD simulation time steps at 300 K for optimized pure SrO (a), B-SrO (b), N-SrO (c), BN-SrO-I (d), and BN-SrO-II (e). The blue solid line is the variation of total energy with time for all considered monolayers.}
	\label{fig07}
\end{figure}

The thermal stability of a pure and BN-codped SrO is tested for approximately $5$ ps with a time step of $1.0$~fs as is presented in \fig{fig07}. The temperature curve of the pure and the doped SrO monolayers neither displays large fluctuations in the temperature nor serious structure disruptions or bond breaking at $300$~K. This indicates that the pure and the doped SrO monolayers are thermodynamically stable structures. In addition, The variation of total
energy (blue solid line) per atom is less than 1 eV which is in the acceptable range similar to many studies in the literature \cite{D1CP01183A}.

\begin{figure}[htb]
	\centering
	\includegraphics[width=0.45\textwidth]{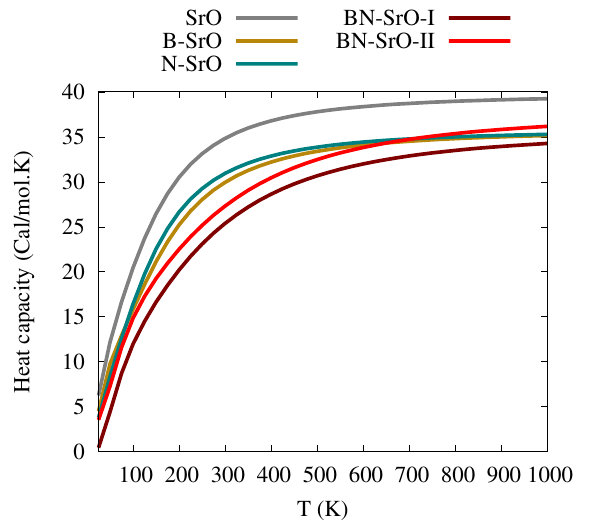}
	\caption{Heat capacity versus temperature of pure SrO (gray), B-SrO (gold), N-SrO (teal), BN-SrO-I (maroon), and BN-SrO-II (red).}
	\label{fig08}
\end{figure}

The inter-atomic interactions influence the thermal properties, such as the heat capacity as is shown in \fig{fig08}. The heat capacity is calculated from $50$ K to the high temperature of $1000$~K, where the heat capacity indicates the ratio of heat absorbed by the material to its temperature change.
The heat capacity is enhanced with temperature and it approaches a constant from
$T>400$~K for undoped and doped SrO monolayers. 
The heat capacity at room temperature is $\approx 35$ for pure SrO monolayer, and 
it is decreased in the cases of doped SrO. 

The predicted trend for the heat capacity of the monolayers is consistent with
the classical theory, that expects higher heat capacity for the systems with stronger
bonds \cite{MORTAZAVI2021100257}. A strong bond is achieved if the electronegativity difference across a bond is high. The average electronegativity difference across the bonds is $2.49$ (SrO), $2.33$ (N-SrO), $2.14$ (B-SrO), $2.11$ (BN-SrO-II), and $2.06$ (BN-SrO-I).

One can see that the average electronegativity across the Sr-O bond of a pure SrO is low among all considered monolayers. This leads to the highest heat capacity of the pure SrO monolayer, and the heat capacity is gradually decreased for the corresponding doped SrO consistent to the value of average electronegativity of the monolayers.

\subsection{Optical properties}

The optical response of pure and BN-codoped SrO monolayers to electromagnetic radiation 
are studied here using the random phase approximation (RPA).
The optical spectral parameters can be obtained from the real part, Re($\varepsilon$) or $\varepsilon_1$, and the imaginary part, Im($\varepsilon$) or $\varepsilon_2$, of the dielectric function, which are collected in $\varepsilon(\omega) = \varepsilon_1(\omega) + i \varepsilon_2(\omega)$. The absorption coefficient, $\alpha$, and the optical conductivity, $\sigma_{\rm optical}$, can be calculated from both parts of the dielectric function \cite{ABDULLAH2023107163, ABDULLAH2023140235}. Since the N-SrO and B-SrO have a metallic property, both the intraband and the interband transitions are taken into account for these two monolayers, and only the interband transitions are calculated for the pure SrO, BN-codped SrO monolayers which have a semiconductor character \cite{RAI2014355}.

\begin{figure}[htb]
	\centering
	\includegraphics[width=0.5\textwidth]{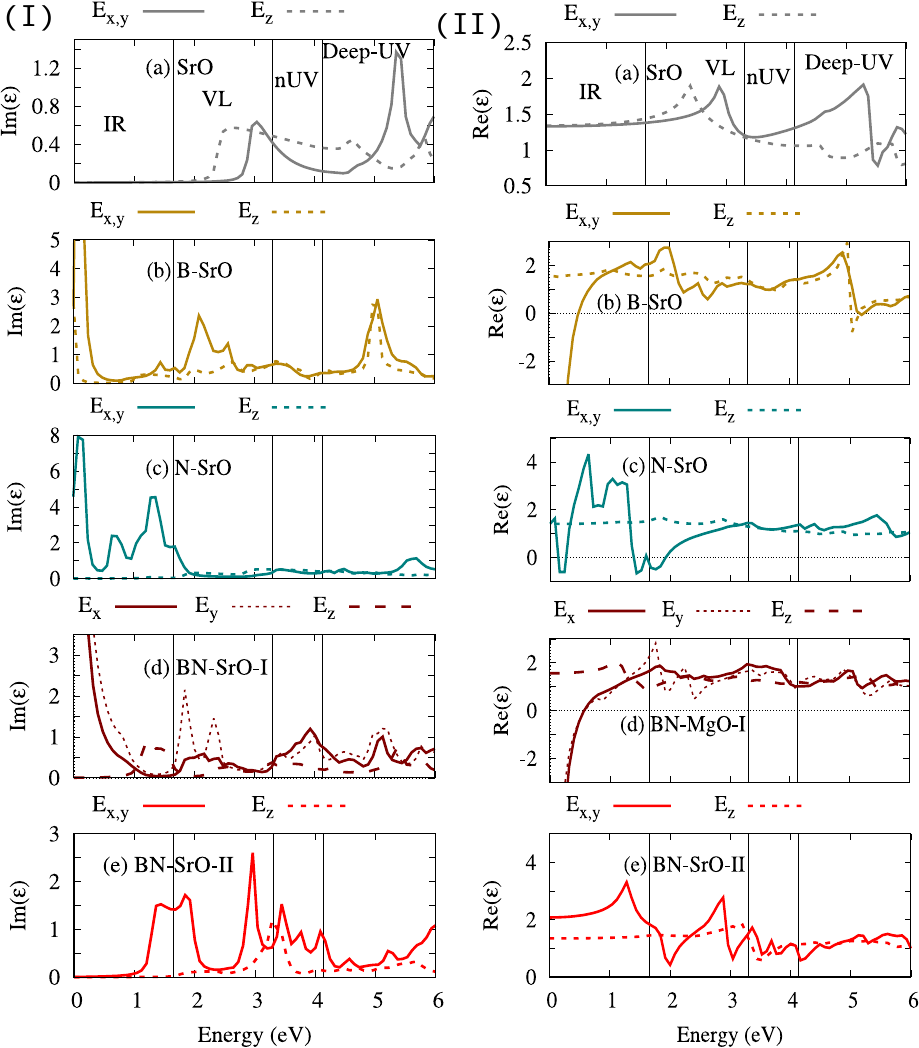}
	\caption{The Im$(\varepsilon)$ (I) and the Re$(\varepsilon)$ (II) presented for pure and BN-codoped SrO monolayers. The solid-, dotted-, and dashed curves correspond to
    the incident light polarized along the $x$, $y$, and $z$ directions, respectively.}
	\label{fig09}
\end{figure}

The Im($\varepsilon$) (I) and Re($\varepsilon$) (II) are shown in \fig{fig09}. 
We first check the isotropic property of the monolayers from the dielectric functions, and 
one can see that all the considered structure, except BN-SrO-I, have an isotropic behavior along the $x$- and the $y$-axis. The anisotropy of BN-SrO-I can be referred to the strong attractive interaction between the B and the N atoms, that makes the BN-SrO-I to have a small asymmetric shape along the $x$- and the $y$-axis.  
Consequently, the optical responses along these two directions will be different.

It is known that the peaks in the Im($\varepsilon$) are caused by the absorption of incident photons
indicating interband transitions of electrons, but the intense peak of both B-SrO (I-b) and N-SrO (I-c) in IR region is formed due to intraband transition confirming the metallic property of these two monolayer.
We see that the highest peaks of Im($\varepsilon$) for E$_x$ are at the photon energy of $3.05$, $0.09$, and $1.36$ eV for pure SrO, BN-SrO-I, and BN-SrO-II, respectively, confirming the semiconductor property of these monolayers. 
Similarly, Re($\varepsilon$) shows maximum peaks at $2.88$,
$1.85$ and $1.28$~eV in E$_x$ direction for pure SrO, BN-SrO-I, and BN-SrO-II, respectively. These
values correspond to the interband transitions between the CBM and
the VBM states and are consistent with the electronic density of states
as given in \fig{fig04} and \fig{fig05}.

In addition, the localized states appearing in the DOS of BN-SrO-I in the conduction band region (see \fig{fig05}(b)) lead to localized peaks in Im($\varepsilon$) in the E$_y$ direction. The localized peaks indicates more transition channels due to the strong attractive interaction between the B and the N atoms.

\begin{figure}[htb]
	\centering
	\includegraphics[width=0.5\textwidth]{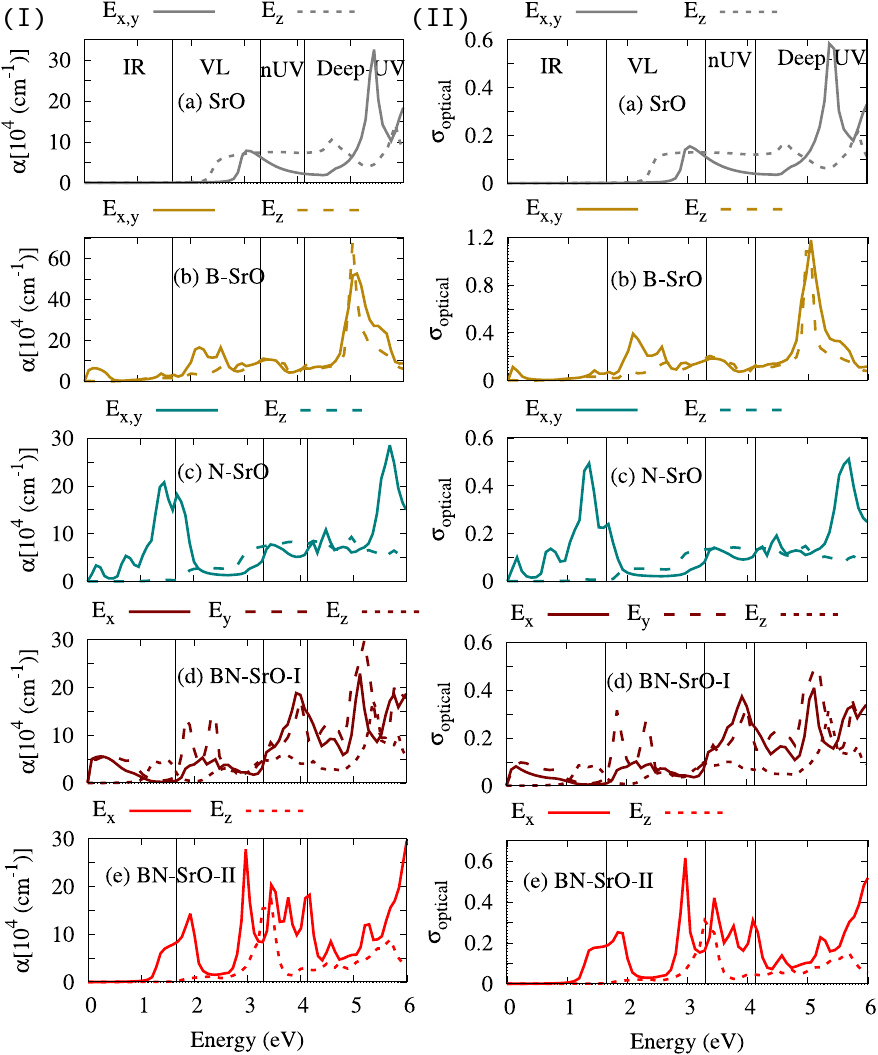}
	\caption{The absorption coefficient, $\alpha$ (I), and real part of optical conductivity, $\sigma_{\rm optical}$ (II) of the pure and BN-codoped SrO monolayers.}
	\label{fig10}
\end{figure}

The absorption coefficient (I) and optical conductivity (II) for pure and BN-codoped SrO monolayers are presented in \fig{fig10} are directly related to the energy band structure of these materials. The $\alpha$ and $\sigma_{\rm optical}$ have almost the same qualitative character, we thus describe them together.

The optical conductivity is increased from low to higher photon energies and the related peaks 
show high conductivity by the photons in these energy ranges presenting the regions of deeper
penetration for electromagnetic waves. 
The highest peaks of real optical conductivity of pure SrO monolayer are obtained at Deep-UV region for both E$_x$ and E$_z$ directions. The intense peak representing the transitions from the CBM to the VBM is found in the VL or IR region which can determine the optical band gap of the monolayer. For pure SrO, the peak representing the optical band gap is seen at $3.04$~eV for E$_x$ and $2.6$~eV for E$_z$ direction which are larger than the electronic indirect band gap found in \fig{fig03}(a).

Similarly, a peak in VL region is seen for BN-SrO-II at $1.85$~eV for E$_x$ direction corresponding to the optical band gap which is again larger than the electronic indirect band gap observed in \fig{fig03}(e). But the first peak for BN-SrO-I is found at $0.15$~eV in which the optical band gap is equal to the electronic band gap because the monolayer has a direct band gap. Interestingly, two pronounced peaks in optical conductivity and absorption coefficient of BN-SrO-I are clearly noticed in the VL region which are caused by the localized peaks in the DOS formed due to the inter-atomic attractive interactions between dopant atoms. We can thus confirm that the BN-codoped SrO can operate well in the visible light region.

\section{Conclusion}\label{conclusion}

The first principle study of pure and BN-codoped SrO monolayer was demonstrated in this work. 
Motivated by the latest advancing theoretical work in the design and fabrication of graphene-like
SrO monolayer, the electronic, the thermal, the magnetic, and the optical characteristics were investigated. 
A substantial reduction of the electronic band gap was found when a strong attractive interaction between the BN-dopant atoms is present in the system. One can thus remarkably see an enhancement in the optical behavior, reduction in thermal property, and the magnetic phase change due to interaction effects between the B and the N atoms.
These results suggest an appealing character of the SrO monolayer for the design of thermal, management, and optical systems.

\section{Acknowledgment}
This work was financially supported by the University of Sulaimani and
the Research center of Komar University of Science and Technology.
The computations were performed on resources provided by the Division of Computational
Nanoscience at the University of Sulaimani.

%\section{References}

%\bibliographystyle{elsarticle-num}
%\bibliography{Ref_2.bib}

\end{document}